\documentclass[times,english]{elsarticle}
\usepackage{color}
\usepackage[colorlinks=true]{hyperref}
\usepackage{float}
\usepackage{stmaryrd}
\usepackage{babel}
\usepackage{units}
\usepackage{textcomp}
\usepackage{textcomp}
\usepackage{amsmath}
\usepackage{commath}
\usepackage{amssymb}
\usepackage{graphicx}
\usepackage{esint}
\usepackage{caption}
\usepackage{subcaption}
\usepackage{epstopdf}

\biboptions{sort&compress}

\makeatletter
\journal{Chaos, Solitons \& Fractals}

\begin{document}

\begin{frontmatter}

\title{A quasi-periodic route to chaos in a parametrically driven nonlinear medium}
\author[UTA1]{Ana M. Cabanas\corref{mycorrespondingauthor1}}
\cortext[mycorrespondingauthor1]{Corresponding author}
\ead{ana.cabanas.plana@gmail.com}
\author[UTA2,UNEFM]{Ronald Rivas}
\author[UTA2]{Laura M. P\'erez}
\author[UTA2,CEDENNA]{Javier A. V\'elez}
\author[UFRO]{Pablo D\'iaz}
\author[UCHILE]{Marcel G. Clerc}
\author[MPIP]{Harald Pleiner}
\author[UTA2]{David  Laroze\corref{mycorrespondingauthor2}}
\cortext[mycorrespondingauthor2]{Corresponding author}
\ead{dlarozen@uta.cl}
\author[TelAviv]{Boris A. Malomed}

\address[UTA1]{Sede Esmeralda, Universidad de Tarapac\'{a}, Av. Luis Emilio Recabarren 2477, Iquique, Chile}
\address[UTA2]{Instituto de Alta Investigaci\'{o}n, Universidad de Tarapac\'{a}, Casilla 7D, Arica, Chile}
\address[UNEFM]{Departamento de F\'{i}sica, Universidad Nacional Experimental Francisco de Miranda, Coro, Venezuela}
\address[CEDENNA] {Centro de Nanociencia y Nanotecnología, CEDENNA, Av. Libertador Bernardo O'Higgins 3363, Santiago, Chile}
\address[UFRO]{Departamento de Ciencias F\'{i}sicas, Universidad de La Frontera, Casilla 54-D, Temuco, Chile}
\address[UCHILE]{Departamento de F\'isica  and Millennium Institute for Research in Optics, Facultad de Ciencias F\'isicas y Matem\'aticas, Universidad de Chile, Casilla 4873, Santiago, Chile}
\address[MPIP]{Max Planck Institute for Polymer Research, D 55021 Mainz, Germany}
\address[TelAviv]{Department of Physical Electronics, School of Electrical Engineering, Faculty of Engineering, Tel Aviv University, Tel Aviv 69978, Israel}

\begin{abstract}
Small-sized systems exhibit a finite number of routes to chaos. However, in extended systems,
not all routes to complex spatiotemporal behavior have been fully explored.
Starting from the sine-Gordon model of parametrically driven chain of damped nonlinear oscillators,
we investigate a route to spatiotemporal chaos emerging from standing waves.
The route from the stationary to the chaotic state proceeds through quasi-periodic dynamics.
The standing wave undergoes the onset of oscillatory instability, which subsequently exhibits a different critical
frequency, from which the complexity originates.
A suitable amplitude equation, valid close to the parametric resonance, makes it possible to produce universe results.
The respective phase-space structure and bifurcation diagrams are produced in a numerical form.
We characterize the relevant dynamical regimes by means of the largest Lyapunov exponent,
the power spectrum, and the evolution of the total intensity of the wave field.
\end{abstract}
\end{frontmatter}

%%%%%%%%%%%%%%%%%%%%%%%%%%%%%%%%%%%
%%%%%%%%%%%%%%%%%%%%%%%%%%%%%%%%%%%

\section{Introduction}

\label{intro}

A natural way to pump energy into various dynamical systems is offered by
the use of direct or parametric resonances \cite{Landau1976}. Direct
resonances were exploited at the dawn of modern physics by Galileo \cite{Galileo} in the characterization of a pendulum. Parametric resonances were
highlighted by the pioneering works of Faraday \cite{Faraday1831} in the
study of vibration modes of a water tank. Theoretical and experimental
studies of parametrically forced nonlinear systems are relevant to many
settings in physics, chemistry, biology, and engineering. Due to the
competition between injection and dissipation of energy, the driven systems
demonstrate complex behavior, leading to the formation of a wide range of
spatiotemporal structures, such as localized ones, extended patterns, and
chaotic states, among others \cite%
{Nic77,Haken1977,CrossHohenberg1993,pismen2006,Manneville2004,Walgraef2012,Murray,Nicolis1995}. One of the most generic scenarios of the self-organizing behavior is
pattern formation. With the increase of the energy-injection rate, basic
stationary patterns may become unstable and bifurcate into a great variety
of more complex ones \cite%
{CrossHohenberg1993,pismen2006,Nicolis1995,CoulletDefectmediateA,CoulletDefectmediateB,ClercVerschueren2013,BrandA,BrandB,BrandC,BrandD,BrandE,Descalzi16,Descalzi18,Descalzi20}. In particular, a manifestation of the complex spatiotemporal dynamics is
time-aperiodic behavior, as observed, \textit{inter alia}, in fluids \cite{DeckerA,DeckerB,DeckerC, Miranda, Brunet}, chemical waves \cite{Flesselles}, cardiac fibrillation \cite{Weiss1997}, granular matter \cite{Swinney2001},
electroconvection \cite{Ahlers2006}, and photonic devices \cite%
{Verschueren2013,ClercGonzalezWilson20016,ClercGonzalezAlvarez2020}. In the
broad variety of physical setups exhibiting complex spatiotemporal
behaviors, routes leading to the emergence of spatiotemporal chaos are not
fully understood yet, as analytical investigation of the underlying
nonlinear partial differential equations is a challenging problem. Diverse
routes have been identified for the transition from order to disorder in
wave settings, such as defect-mediated turbulence \cite%
{CoulletLega1988,CoulletLegaGil1988A,CoulletLegaGil1988B}, onset of
spatiotemporal chaos in Galilean-invariant systems \cite%
{GorenEckmannProcaccia1998}, spatiotemporal intermittence \cite{Chate1994},
quasi-periodicity of travelling \cite{Brochard2006} and standing waves \cite%
{ClercVerschueren2013}, crisis of spatiotemporally chaotic saddles \cite%
{He1998,Chian2004,Chian2007}, spatiotemporal chaos which sets in via
stationary branching shocks and holes \cite{Sherratt2012}, and the phase
turbulence \cite{Kuramoto1984}.

A generic model that describes periodically forced systems is based on the
parametrically driven damped nonlinear Schr\"{o}dinger (PDDNLS) equation
\cite{Miles1984, Barashenkov91,book}. This amplitude equation produces a
variety of temporal behaviors, including stationary, periodic, and chaotic
regimes, such as Faraday waves \cite{Faraday1831,Coullet94,Scott,CCL10},
single-soliton \cite{Barashenkov91,Barashenkov11a,Alexeeva00,Barashenkov99,Zemlyanaya} and
two-soliton \cite{Barashenkov11b,Urzagasti12} states, and spatiotemporal
chaos \cite{Barashenkov91,Shchesnovich02}. Remarkable hydrodynamic modes are
excited by the parametric instability in the form of standing (Faraday)
waves on the surface of a vertically vibrated Newtonian fluid \cite%
{Faraday1831}. These standing waves respond strongest at a half of the
forcing frequency (the $2:1$ resonance) \cite{Arnold}. The basic PDDNLS
equation describes parametric instabilities close to the $2:1$ resonance in
weakly dissipative systems \cite{CCL09b}. However, this equation in its
simplest form is not able to produce certain phenomena, such as stable
localized modes that connect a uniform oscillatory state to an extended wave
\cite{CCL12}, or domain walls between uniform oscillations with opposite
phases \cite{CCL10b,CCL09b,CCL08,CCL09a}. To study the existence and
properties of a broad class of dynamical states, a generalized PDDNLS
equation, produced by a systematic expansion procedure, can be used. It
adequately accounts for diverse types of the dynamical behavior, such as
fronts and confined patterns \cite{Urzagasti14}.

This work aims to investigate dynamical scenarios based on standing waves in
a parametrically driven chain of pendula, which is considered in the
continuum approximation. Starting from the respective parametrically-driven
damped sine-Gordon model, it is shown that spatiotemporal chaos can emerge
from standing waves. The chaos appears as an extension of quasi-periodic
dynamics. Namely, changing parameters of the model, we observe that the
standing wave exhibits an oscillatory instability. Further variation of the
parameters leads to a secondary oscillatory instability, which eventually
gives rise to the spatiotemporal chaos. The generalized PDDNLS equation,
valid close to the parametric resonance, allows to systematically
investigate the emergence of the complex behavior in a numerical form. The
largest Lyapunov exponent, power spectrum, and time evolution of the total
norm of the wave field are computed to characterize the transitions between
stationary, quasi-periodic, and chaotic dynamical states.

The manuscript is organized as follows. The physical model, the nonlinear
analysis and the dynamical indicators used to describe the system are
presented in Sec. \ref{sec:2}. Results produced by systematic numerical
investigation of the patterns and characterization of the dynamical behavior
of the system are reported in Sec. \ref{sec:3}. Finally, conclusions are
formulated in Sec. \ref{sec:4}. %%%%%%%%%%%%%%%%%%%%%%%%%%%
%%%%%%%%%%%%%%%%%%%%%%%%%%%

\section{The parametrically driven damped sine-Gordon equation and the
nonlinear-Schr\"{o}dinger approximation \label{sec:2}}

\begin{figure}[h]
\begin{center}
\includegraphics[width=\textwidth]{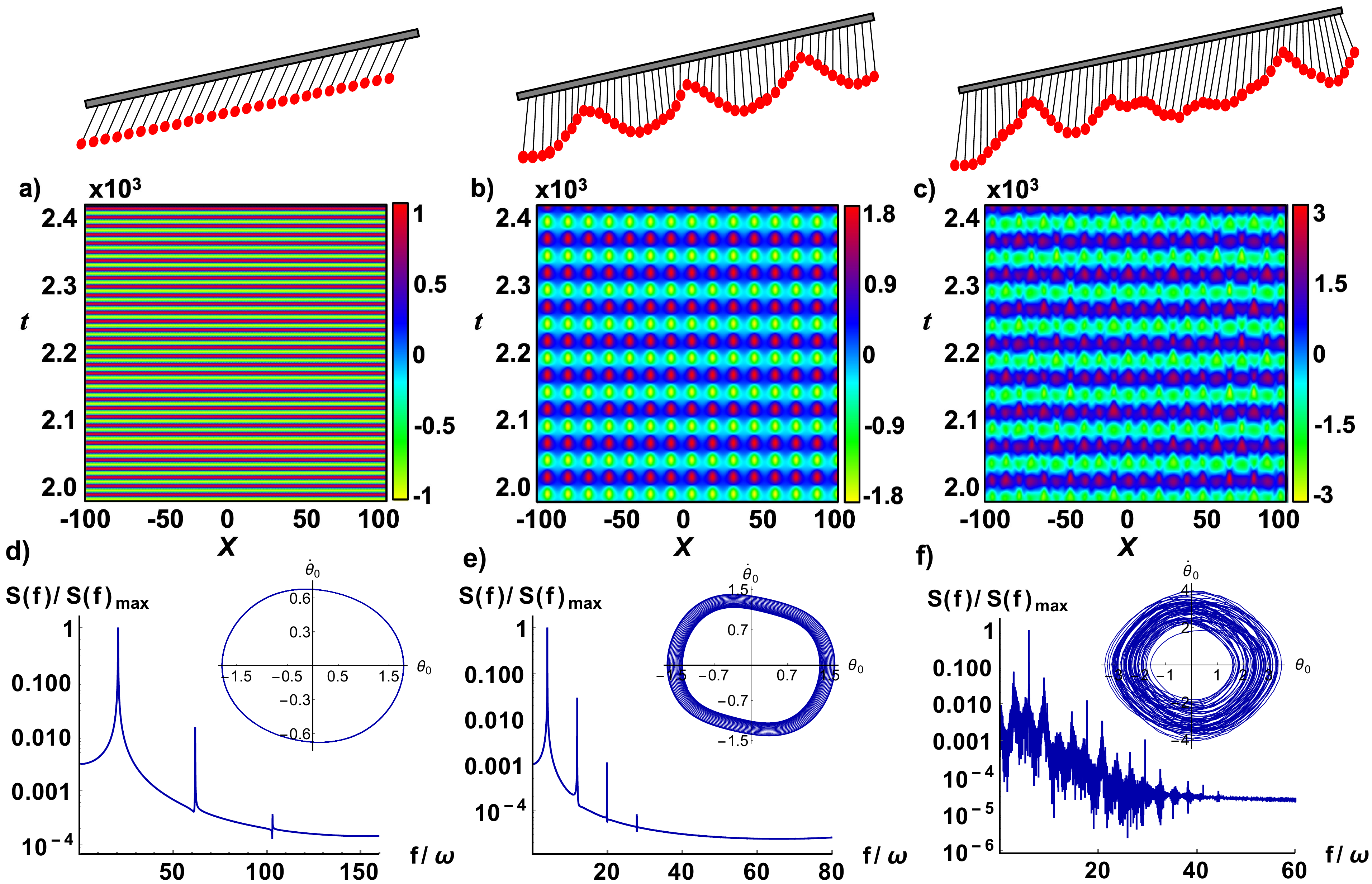}
\end{center}
\caption{Spatiotemporal dynamical behavior produced by the parametrically
driven damped sine-Gordon equation (\protect\ref{Eq-sineGordon}).
Color-coded spatiotemporal distributions of the continuous field, $\protect%
\theta \left( x,t\right) $, are displayed, with time interval $\Delta
\protect\tau =2000$, in the middle row, parallel to typical configurations
of the chain of coupled pendula shown in the top row, which is approximated
by Eq. (\protect\ref{Eq-sineGordon}) in the continuum limit. a) A spatially
uniform state obtained at the amplitude and detuning of the parameteric
drive $\protect\gamma _{0}=0.85$ and $\protect\nu =-0.02$ (see Eq. (\protect
\ref{nu})), with the largest Lyapunov exponent $\protect\lambda _{\max
}=-0.037$. b) A regular spatially periodic pattern at $\protect\gamma %
_{0}=0.40$ and $\protect\nu =-0.05$, with $\protect\lambda _{\max }=-0.041$
c) A chaotic state at $\protect\gamma _{0}=0.61$ and $\protect\nu =-0.08$,
with $\protect\lambda _{\max }=0.055$. Bottom panels d), e), and f) show
normalized Fourier spectra $S(f)$ and phase portraits in the plane of $%
\left( \protect\theta ,\dot{\protect\theta}\right) $ at a fixed spatial
point ($x=0$), plotted for a), b) and c), respectively. Other parameters in
Eq. (\protect\ref{Eq-sineGordon}) are $\protect\omega _{0}=1$, $\protect\mu %
_{0}=0.1$, and $\protect\kappa =1$.}
\label{Fig1}
\end{figure}

We start by consideration of a chain of parametrically driven damped pendula
(an extended Scott's model \cite{Scott}), described in the continuum limit
by the parametrically driven damped sine-Gordon equation \cite{Coullet94}:
\begin{equation}
\ddot{\theta}=-\left[ \omega _{0}^{2}+\gamma _{0}\sin (\omega t)\right] \sin
\theta -\mu _{0}\dot{\theta}+\kappa {\partial _{xx}}\theta ,
\label{Eq-sineGordon}
\end{equation}%
where $\theta (x,t)$ is the angle between the pendulum at position $x$ and
the vertical axes at time $t$, the overdot stands for the time derivative, $%
\omega _{0}$ is the eigenfrequency of the pendulum, $\gamma _{0}$ and $%
\omega $ are the amplitude and frequency of the parametric forcing, $\mu
_{0} $ is the damping coefficient and $\kappa $ the coefficient of the
elastic coupling. In the following, by means of rescaling of $x$ and $t$ we
fix $\kappa =\omega _{0}=1$. Figure~\ref{Fig1} illustrates different types
of dynamical oscillatory states in the system.

%%F

%%F

\begin{figure}[tbp]
\begin{center}
\includegraphics[width=0.7\linewidth]{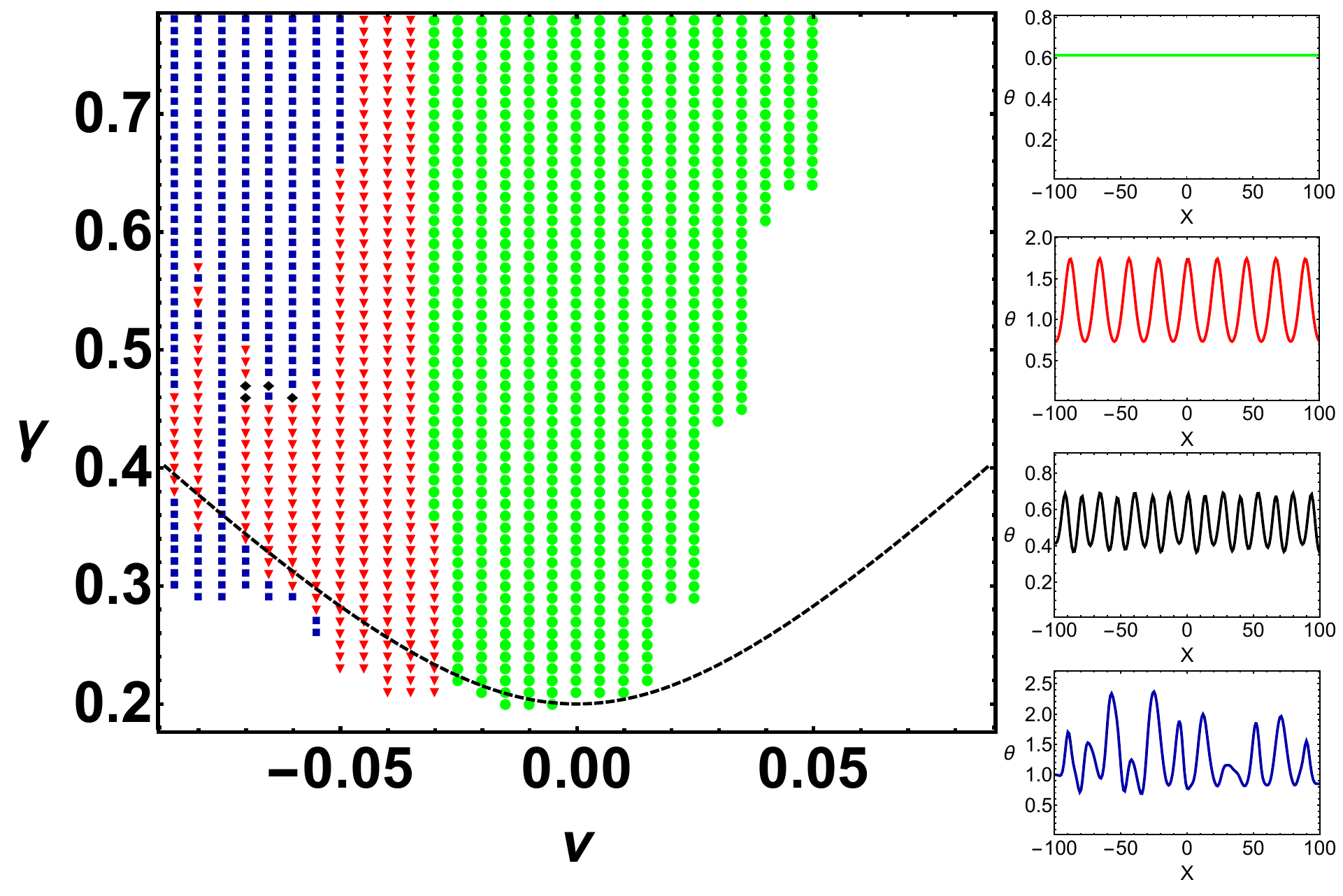}
\end{center}
\caption{The diagram of dynamical states in the $\left( \protect\gamma ,%
\protect\nu \right) $ parameter plane of Eqs. (\protect\ref{Eq-sineGordon})
and (\protect\ref{nu}). The dashed curve represents the instability boundary
(Arnold tongue) given by Eq. (\protect\ref{tongue}). Green circles (${\
\color{green}\bullet }$) represent spatially uniform states, red triangles ({%
\ \color{red} $\blacktriangledown $}) denote standing waves, blue squares ({%
\ \color{blue} $\blacksquare $}) chaotic waves, and black diamonds ({\
\color{black} $\Diamondblack$}) correspond to quasi-periodic states with
small amplitudes. Here and in Fig. \protect\ref{Fig6-ArnoldTongue} below,
the trivial state, $\protect\theta \equiv 0$, occurs in the white area.
Right panels display snapshots of the respective spatial patterns. The fixed
parameters are the same as in Fig. \protect\ref{Fig1}.}
\label{Fig2}
\end{figure}

The trivial solution $\theta (x,t)=0$ can be made unstable by the parametric
forcing with the frequency close to the resonant (double) value,
\begin{equation}
\omega =2(\omega _{0}+\nu )  \label{nu}
\end{equation}%
where $\nu $ is a detuning parameter, with $|\nu |\ll \omega _{0}$. The
commonly known analysis \cite{Landau1976} produces the stability boundary
(the so-called \textit{Arnold tongue}), in the form of
\begin{equation}
\nu ^{2}+(\mu _{0}/2)^{2}=(\gamma _{0}/4)^{2},  \label{tongue}
\end{equation}%
depicted in Fig.~\ref{Fig2} by the dashed black hyperbola in the parameter
plane of the detuning and forcing strength. While crossing the boundary, the
destabilization occurs through a supercritical (subcritical) bifurcation at
positive (negative) values of the detuning \cite{CCL09b}, uniform
oscillations taking place inside of the \textquotedblleft tongue" \cite%
{CCL09a}. The corresponding power spectrum, displayed in panel d) of Fig.~%
\ref{Fig1} shows well-defined peaks determined by the forcing. The
respective inset shows the phase portrait corresponding to a closed limit
cycle.

In the Arnold tongue, the uniform oscillations are observed at small
detuning, as illustrated in Fig.~\ref{Fig2} by green circles. Making the
detuning more negative, the uniform oscillations develop a supercritical
oscillatory spatial instability with a nonzero wavenumber \cite%
{CrossHohenberg1993}. It is characterized by emergence of a novel frequency,
which is not commensurate with the frequency of the uniform oscillations.
The instability gives rise to standing waves, as shown in panels b) and e)
of Fig.~\ref{Fig1}. Note that in the respective phase portrait the limit
cycle transforms into a torus. These standing waves are denoted by red
triangles in the diagram of states displayed in Fig.~\ref{Fig2}. Further
decreasing the detuning, the standing wave develops a complex spatiotemporal
behavior, as illustrated in panels c) and e) of Fig.~\ref{Fig1}. It is seen
that the temporal spectrum at a fixed spatial point now represents dynamical
chaos, and, likewise, the phase portrait corresponds to a strange attractor
\cite{Manneville2004}. Such a complex spatiotemporal behavior, denoted by blue squares
in Fig. \ref{Fig2}, is observed in a broad region of the Arnold tongue.

%%F
%%F

\begin{figure}[h]
\begin{center}
\includegraphics[width=0.6\linewidth]{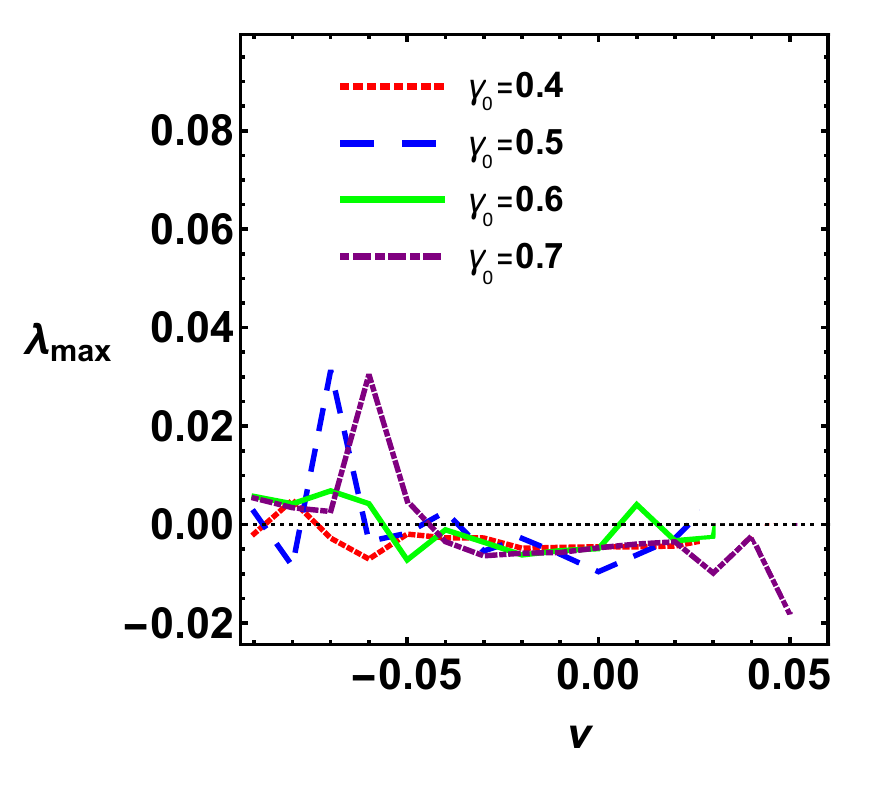}
\end{center}
\caption{The largest Lyapunov exponent $\protect\lambda _{\max }$ vs.
detuning $\protect\nu $ (see Eq. (\protect\ref{nu})), at different values of
the strength of the parametric forcing: $\protect\gamma %
_{0}=(0.4,0.5,0.6,0.7)$. The other parameters are the same as in Fig.
\protect\ref{Fig2}.}
\label{Fig3}
\end{figure}

To characterize the dynamical nature of the complex waves, we use the
largest Lyapunov exponent $\lambda _{\max }$ \cite{Pikovsky2016}. First, $%
\lambda _{\max }<0$ corresponds to a stable equilibrium state, a uniform or
patterned one. On the contrary, $\lambda _{\max }>0$ implies chaotic
dynamics, with exponential sensibility to initial conditions. The case of $%
\lambda _{\max }=0$ corresponds to quasi-periodic behavior or complex
behavior with non-exponential sensitivity. Figure \ref{Fig3} shows $\lambda
_{\max }$ as a function of detuning $\nu $ (see Eq. (\ref{nu})) for
different values of the drive's strength, $\gamma _{0}=\{0.4,0.5,0.6,0.7\}$.
The plots make it possible to distinguish regular and chaotic behavior,
which, as said above, correspond to $\lambda _{\max }<0$ and $\lambda _{\max
}>0$, respectively.

Thus, the parametrically driven damped sine-Gordon equation (\ref%
{Eq-sineGordon}) provides the transition from the spatially uniform
oscillations to the spatiotemporal chaos, via the intermediate spatially
periodic pattern. To achieve a better grasp of these dynamical regimes, in
the next section we develop analysis under conditions of weak nonlinearity
and slowly-varying amplitude, close to the parametric resonance.

%%%%%%%%%%%%%%%%%%%%%%%%%

\subsection{The weak-nonlinearity analysis}

The above model, based in the sine-Gordon equation, can be essentially
simplified for small-amplitude solutions, whose main harmonic frequency is
close to the eigenfrequency, $\omega _{0}$. To this end, the following
ansatz is adopted \cite{CCL09b}:
\begin{equation}
\theta (x,t)=\psi (\tau ,X)e^{-i\omega \tau /2}+\mathrm{c.c.}+\mathrm{h.o.t.},
\label{eq2}
\end{equation}%
where $\psi (\tau ,X)$ is a slowly complex envelope amplitude, $\tau \equiv
t/4,X\equiv x/\sqrt{2}$, while $\mathrm{c.c.}$ and $\mathrm{h.o.t.}$ stand
for the complex conjugate and higher-order terms, respectively. Close to the $%
2:1$ parametric resonance, the frequency is $\omega =2(1+\nu )$, as per Eq. (%
\ref{nu}). The substitution of ansatz (\ref{eq2}) in (\ref{Eq-sineGordon})
yields, in the first approximation, the PDDNLS equation \cite{Barashenkov91}%
, which is the basic model of the driven systems close to the $2:1$
parametric resonance:%
\begin{equation}
\frac{\partial \psi }{\partial \tau }=-i\nu \psi -i|\psi |^{2}\psi -i\frac{%
\partial ^{2}\psi }{\partial X^{2}}-\mu \psi +\gamma \bar{\psi},  \label{eq3}
\end{equation}%
where $\mu \equiv \mu _{0}/2$, $\gamma \equiv \gamma _{0}/4$, , and $\bar{%
\psi}$ is the complex conjugate of $\psi $. This equation is known to
produce stationary, time-periodic, or chaotic solutions, including Faraday
waves \cite{Coullet94,CCL10b}, soliton-like modes \cite%
{Barashenkov91,Barashenkov11a}, two-soliton \cite{Barashenkov11b,Urzagasti12}
and soliton-antisoliton \cite{Urzagasti14b} bound states, and spatiotemporal
chaos \cite{Barashenkov91,Shchesnovich02}.

The PDDNLS equation admits different spatially uniform solutions. The
instability boundary of the trivial one, $\psi =0$, is
\begin{equation}
\gamma ^{2}=\mu ^{2}+\nu ^{2},  \label{tongue2}
\end{equation}%
cf. Eq. (\ref{tongue}). In addition, the trivial solution suffers an
instability which sets in at $\gamma =\mu $ for the positive detuning \cite%
{Coullet94,Elphick87}. Nontrivial uniform solutions of Eq. (\ref{eq3})
are
\begin{equation}
\psi _{\pm ,\pm }=\pm x_{0}(1\pm iy_{0}),  \label{+-}
\end{equation}%
where $x_{0}=\sqrt{(\gamma -\mu )(\phi -\nu )/2\gamma }$ and $y_{0}=\sqrt{%
(\mu -\gamma )/(\mu +\gamma )}$ with $\phi \equiv \sqrt{\gamma ^{2}-\mu ^{2}}
$and two independent signs $\pm $. Note that the uniform solutions emerge at
$\gamma =\mu $ through a saddle-node bifurcation. They are unstable in the
parameter space of Eq. (\ref{eq3}), with the exception of the case of zero
detuning ($\nu =0$), in which case ${\psi }_{\pm ,+}$ is marginally stable
\cite{CCL08}.

In fact, the underlying sine-Gordon equation (\ref{Eq-sineGordon}) may have
stable solutions for uniform oscillations. To provide this possibility, as
well as the existence of more complex states that bifurcate from the uniform
oscillations, it is necessary to proceed to the generalized PDDNLS equation.
Such a generalization can be written as
\begin{equation}
\frac{\partial \psi }{\partial \tau }=-i\nu \psi -i|\psi |^{2}\psi -i\frac{%
\partial ^{2}\psi }{\partial X^{2}}-\mu \psi +\gamma \bar{\psi}+\mathcal{N}%
_{\psi },  \label{eq4}
\end{equation}
with additional terms $\mathcal{N}_{\psi }$ produced by the
weak-nonlinearity expansion of underlying equations of the sine-Gordon type
\cite{CCL09b,CCL09a}. Thus, at order $\gamma ^{5/2}$ of the expansion, one can derive, in
the general case,
\begin{equation}
\mathcal{N}_{\psi }=\gamma \left( b\overline{\psi }|\psi |^{4}+\delta {\psi }%
^{3}+\beta {\psi }^{3}|\psi |^{2}\right) +ia\psi |\psi |^{4}+\eta \frac{%
\partial ^{2}\psi }{\partial X^{2}}+c\psi |\psi |^{2}+i\zeta \overline{\psi }%
\left( \frac{\partial \psi }{\partial X}\right) ^{2},  \label{eq5}
\end{equation}%
where $\eta $ accounts for the diffusion, $c$ represents the nonlinear
dissipation, ($b$, $\delta $, $\beta $) are coefficients of the nonlinear
parametric forcing, $a$ controls the quintic nonlinearity, and $\zeta $ is
the coefficient of the nonlinear-drift term. The amended equation (\ref{eq4} has been employed to produce kinks, localized patterns and
traveling pulses \cite
{CCL10b,CCL09b,CCL12,CCL08,CCL09a,Urzagasti14,Urzagasti14b,CCL10,Urzagasti17a,Elphick87,Malomed2009,Bara2003,KnoblochA,KnoblochB,LECS15,Edri20,Edri220}%
. In fact, not all the additional terms are necessary to secure the
stability of the uniform states inside of the Arnold tongue \cite{Bara2003}.
Note that, in the case of the sine-Gordon equation, the expansion at order $%
\gamma ^{5/2}$ yields $\eta =0$, $c=0$, $\alpha =1/2$, $\delta =-1/6$, $%
a=1/6 $, $b=1/12$, $\beta =-1/24$, and $\zeta =0$ \cite{Urzagasti14}.

We here focus on the case of $\eta =0$ and $c=0$, as effects of the terms $%
\sim \eta $ and $c$ have been studied before \cite%
{Bara2003,KnoblochA,KnoblochB}. Besides, the effect of $\zeta $ is
neglected, as it is known too, inducing traveling solutions \cite{LECS15}.
Then, it is easy to find that the squared amplitude of the uniform-state
solution, $|\psi |^{2}$, is determined by equation

\begin{equation}
\gamma ^{2}=\frac{\mu ^{2}}{\left[ 1+\left( \delta +\alpha \right) \,|\psi
|^{2}+\left( \beta +b\right) \,|\psi |^{4}\right] ^{2}}+\frac{(\nu +|\psi
|^{2}-a\,|\psi |^{4})^{2}}{\left[ 1+\left( \alpha -\delta \right) \,|\psi
|^{2}+\left( b-\beta \right) \,|\psi |^{4}\right] ^{2}}.  \label{eq6}
\end{equation}

To study small-amplitude patterns generated by the extended PDDNLS equation,
the coefficients of $\mathcal{N}_{\psi }$ in Eq. (\ref{eq5}) are introduced
as free parameters. In particular, numerical results are displayed below for
$b=1/12$, $\delta =4/15$, $\beta =-1/24$, $\alpha =-0.65$, $a=1/6$, and $\mu
=0.275$, which make it possible to produce generic findings.

\subsection{Indicators of complexity and transition between dynamical regimes}

%%F
\begin{figure}[tbp]
\begin{center}
\includegraphics[width=\textwidth]{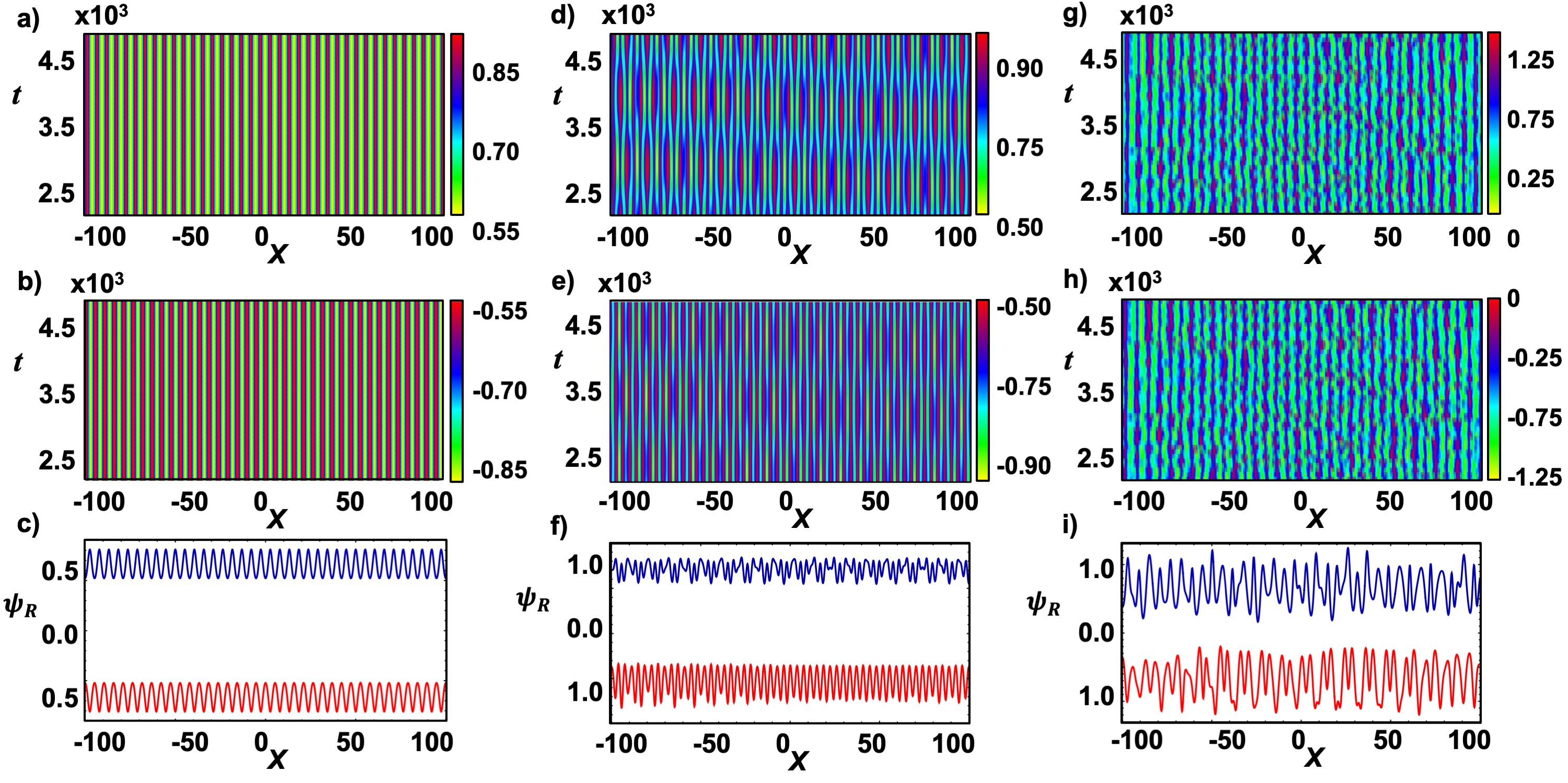}
\end{center}
\caption{The spatiotemporal evolution of the real part of $\protect\psi $ in
established dynamical states shown within a time interval $\Delta \protect\tau %
=2000$ (the first and second rows), and snapshots of the respective spatial
patterns at a certain time (the third row). a), b) and c): Two regular
small-amplitude patterns ${\protect\psi }_{+/-}$ (in- and out-of-phase ones,
see the text) at $\protect\gamma =0.85$ and $\protect\nu =-0.25$, with $%
\protect\lambda _{\max +}=-1.853\times 10^{-3}$ and $\protect\lambda _{\max
-}=-1.852\times 10^{-3}$. d), e) and f): Quasi-periodic states ${\protect%
\psi }_{+/-}$ at $\protect\gamma =0.85$ and $\protect\nu =-0.45$, with $%
\protect\lambda _{\max +}=4.92\times 10^{-4}$ and $\protect\lambda _{\max
-}=5.13\times 10^{-4}$. g), h) and i): Small-amplitude chaotic patterns ${%
\protect\psi }_{+/-}$ at $\protect\gamma =0.85$ and $\protect\nu =-0.55$,
with $\protect\lambda _{\max +}=1.288\times 10^{-2}$ and $\protect\lambda %
_{\max -}=1.359\times 10^{-2}$. Fixed parameters are $b=1/12$, $\protect%
\delta =4/15$, $\protect\beta =-1/24$, $\protect\alpha =-0.65$, $a=1/6$ and $%
\protect\mu =0.275$.}
\label{Fig4}
\end{figure}

To characterize the dynamics produced by Eq. (\ref{eq4}), we use standard
indicators, one of which is the time-dependent energy. It is often used to
study non-regular dynamics in fluids, optics \cite%
{AkhmedievA,AkhmedievB,AkhmedievC,AkhmedievD,Edri20,Edri220} and other physical systems
\cite{Knobloch0607A,Knobloch0607B,Knobloch0607C}:
\begin{equation}
Q(\tau )=\frac{1}{2L}\int_{-L}^{+L}|\psi (\tau ,X)|^{2}\,dX,  \label{eq7}
\end{equation}%
where $2L$ is the system's size. The simple uniform stationary regime has $%
Q= $ $\mathrm{const}$, and it is a (quasi-)periodic function of time in the
case of (quasi-)periodic dynamics. An apparently random time dependence of $%
Q(\tau )$ corresponds to chaotic behavior.

To categorize different dynamical regimes, we also use the power spectrum of
the amplitude, $S_{Q}\left( f\right) =|\mathfrak{F}(f)|^{2}$, where $%
\mathfrak{F}$ is the Fourier transform of $Q(\tau )$,
\begin{equation}
\mathfrak{F}(f)=\frac{1}{\sqrt{2\pi }}\int_{0}^{\tau _{\max }}Q(\tau )\exp
\left( -if\tau \right) d\tau .
\end{equation}%
If the time series is regular, function $S_{Q}\left( f\right) $ has a finite
number of quasi-discrete peaks. On the contrary, the spectrum is essentially
continuous if the series is chaotic \cite{Ott}. In addition, to characterize
spatial fluctuations of field $\psi $, we also use the time average of the
local density,
\begin{equation}
\mathcal{L}(x)=\frac{1}{T-T_{0}}\int_{T_{0}}^{T}|\psi (\tau ,X)|^{2}\,d\tau ,
\label{eq7-2}
\end{equation}%
and its spatial power spectrum,
\begin{equation}
S_{\mathcal{L}}\left( k\right) =\int_{-L}^{+L}S_{\mathcal{L}}\left( k\right)
\exp \left(- ikX\right) \,dX,  \label{S}
\end{equation}%
at a set of points in the frequency space $k=\left( {k_{1},...,k_{n}}\right)
$. Similar to the temporal power spectrum, its spatial counterpart, $S_{%
\mathcal{L}}\left( k\right) $, helps to quantify different types of the
spatial behavior of the system.

Further, a well-known potent indicator of the dynamics is the largest
Lyapunov exponent \cite
{Wolf85,Sprott2003,Pikovsky2016,Mag12A,Mag12B,Mag12C,Eckmann86,Geist90,Cross06A,Cross06B,Gallas10A,Gallas10B,Gallas10C,Velez20,LarPle2015,LarSiddPle2013,Mahmud,Cabanas19,Cabanas18,Perez15,Gallas2020,Cabanas21,SKL2021}%
\begin{equation}
\lambda _{\max }=\underset{\tau \rightarrow \infty }{\text{lim}}\Big[\frac{1%
}{\tau }\;\text{ln}\Big(\frac{\Vert \delta \psi (\tau ,x)\Vert }{\Vert
\delta \psi _{0}\Vert }\Big)\Big],  \label{eq8}
\end{equation}
where $\Vert \bullet \Vert $ stands for the quadratic norm, and $\delta \psi
(\tau ,x)$ is a numerically generated solution of the linearized equation, $%
\partial \left( \delta \psi \right) /\partial \tau =\bar{\mathbf{J}}\cdot
\delta \psi $, with $\bar{\mathbf{J}}$ being he Jacobian matrix of Eq. (\ref%
{eq4}). We remark that $\lambda _{\max }$ predicts how fast
distance $\delta \psi $ between two initially close trajectories of field $%
\psi $ grows in the course of the evolution. Namely, if the system is
chaotic with $\lambda _{\max }>0$, two configurations, that were close at $%
\tau =\tau _{0}$, separate in phase space exponentially fast. On the other
hand, in the case of $\lambda _{\max }<0$ they converge to a stationary
attractor. The marginal case, $\lambda _{\max }=0$, corresponds to
time-periodic, quasi-periodic, or complex dynamics without exponential
sensibility to the variation of initial conditions.

\begin{figure}[tbp]
\begin{center}
\includegraphics[width= \textwidth]{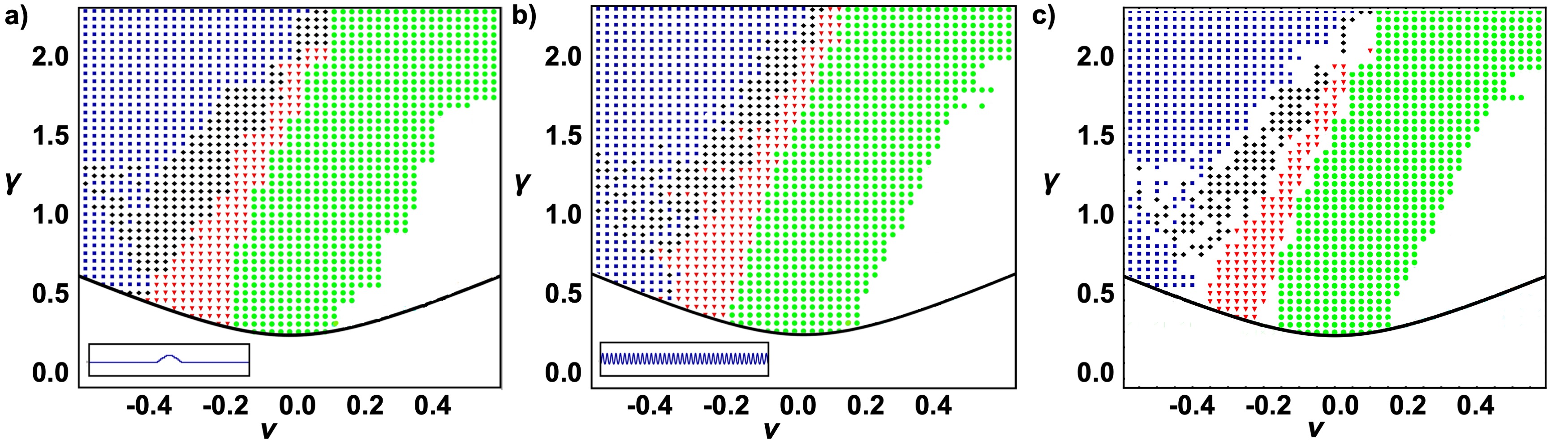}
\end{center}
\caption{Phase diagrams in the $\left( \protect\gamma ,\protect\nu \right) $
parameter plane for different initial conditions shown in the insets. a): A
flat profile with a bump; b) a spatially-periodic profile. Green circles (${%
\ \color{green}\bullet }$) correspond to the spatially uniform state, red
triangles ({\color{red} $\blacktriangledown $}) denote stationary
small-amplitude states, blue squares ({\color{blue} $\blacksquare $}) the
small-amplitude chaotic pattern, and black diamonds ({\color{black} $
\Diamondblack$}) correspond to quasi-periodic states with a small amplitude.
c) Coincidence regions, in which the different initial conditions, used in
diagrams a) and b), produce identical final states. The fixed parameters are
the same as in Fig. \protect\ref{Fig4}.}
\label{Fig6-ArnoldTongue}
\end{figure}

\begin{figure}[h]
\begin{center}
\includegraphics[width=\textwidth]{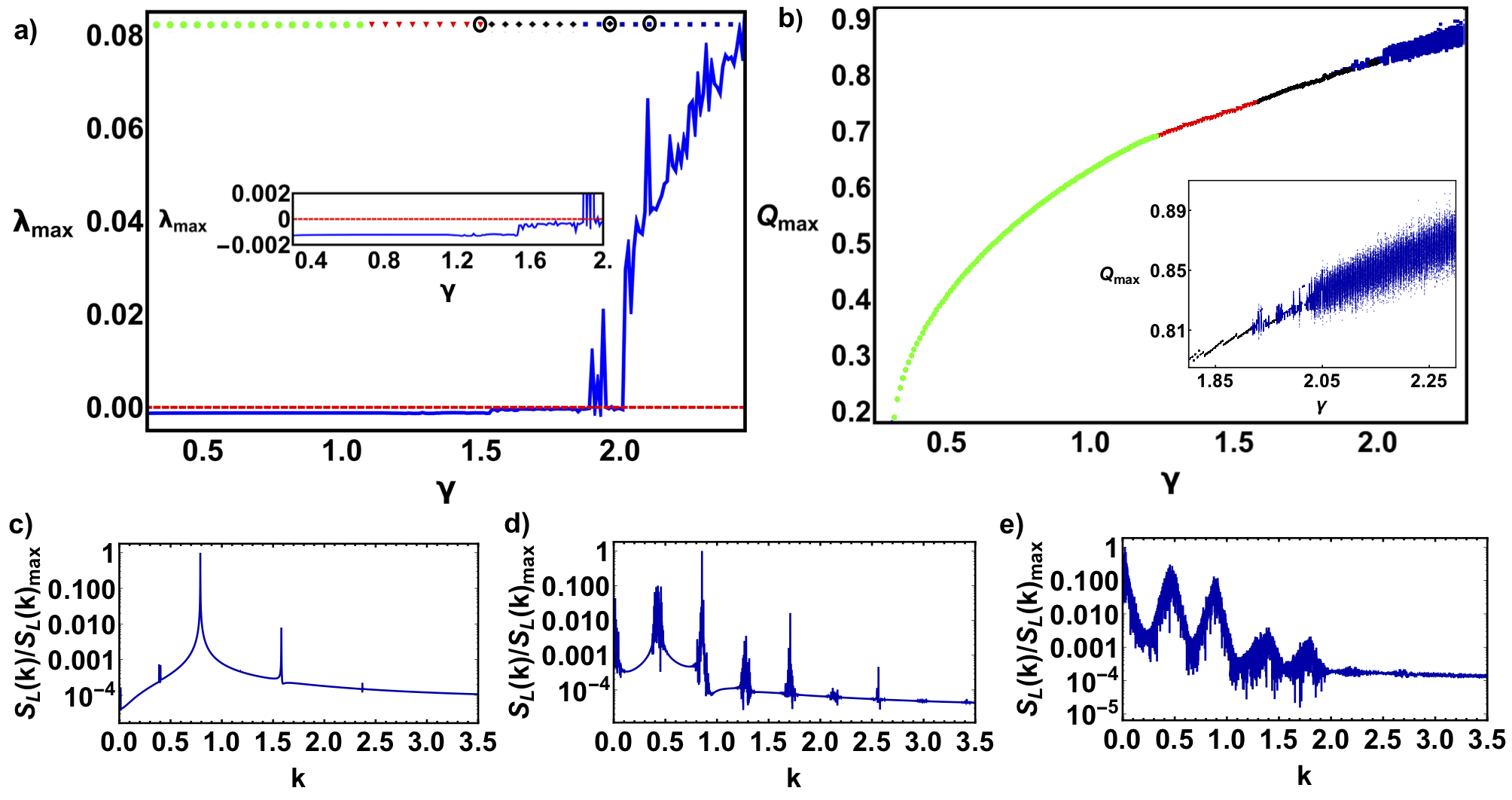}
\end{center}
\caption{Characterization of the route to spatiotemporal chaos produced by
simulations of Eq.~(\protect\ref{eq4}), is shown by means of dynamical
indicators. a) $\protect\lambda _{\max }$, for $\protect\nu =-0.1$, and b) $%
Q_{\max }$, for $\protect\nu =+0.1$, as a function of the forcing strength, $%
\protect\gamma $, see Eqs. (\protect\ref{eq7}) and (\protect\ref{eq8}). The
insets are zoomed-in versions of the main plots. The meaning of the symbols
is the same as in Fig.~\protect\ref{Fig6-ArnoldTongue}: green circles
correspond to spatially uniform states, red triangles to stationary
small-amplitude ones, and blue squares to the small-amplitude chaotic
pattern. Panel c) shows the spatial Fourier spectrum $S_{\mathcal{L}}(k)$
(see Eq. (\protect\ref{S})) for a regular state at $\protect\gamma =1.50$,
d) a temporarily quasi-periodic state at $\protect\gamma =1.95$, and e) a
chaotic one at $\protect\gamma =2.1$.}
\label{Fig6}
\end{figure}

\section{The quasi-periodic route to spatiotemporal chaos in the generalized
PDDNLS equation}

\label{sec:3}

%%%%%%%%%%%%%%%%%%%%%%%%%%%%%

\subsection{Numerically obtained patterns}

A variable-step fifth-order Runge-Kutta scheme \cite{Press92} has been used
for simulations of Eq. (\ref{eq4}) with the precision of $10^{-7}$. The
spatial derivatives were approximated by means of the second-order central
finite-difference method, and the Neumann's boundary conditions were used.
Typical the half-size $L=100$ and $N=1500$ lattice points imply that the
mesh size of the numerical scheme is $\Delta x=200/1500\approx 0.13$.
Sensitivity of the findings to the discretization has been verified by
changing $N$ and $\Delta x$. Variation of the domain size $L$ was used to
verify that finite-size effects do not affect the results. We have used two
essentially different initial conditions, as shown below. The integration
was performed over sufficiently long times (up to $\tau =4.8\times 10^{3}$,
to check that transients have faded out.

For the analysis of spatial patterns we present in Fig.~\ref{Fig4} the
spatiotemporal evolution of the real part of the wave field, $\psi _{\mathrm{%
R}}\equiv $\textrm{Re}$\{\psi \left( x,\tau \right) \}$. Starting close to
the tip of the Arnold tongue, which implies the consideration of a small
detuning ($\gamma =0.85$) and the forcing strength slightly greater than the
dissipation ($\nu =-0.25$, see Eq. (\ref{tongue2})). The regular in-phase and
out-of-phase states ($\psi _{\pm }$, with $\psi _{\mathrm{R}}>0$ and $\psi _{%
\mathrm{R}}<0$, respectively) are observed in Figs.~\ref{Fig4} a)-c). In these panels the
spatial periodicity observed corresponds to the real part of
$\exp \left( ikx\right) $ with some $k$, while $\left\vert \psi \right\vert $
remains constant. The values of $\lambda _{\max }$ for these regular
patterns are negative: $\lambda _{\max +}=-1.853\times 10^{-3}$ and $\lambda
_{\max -}=-1.852\times 10^{-3}$, respectively. 

Following the decrease of the detuning and/or the increase of the forcing
strength, the pattern exhibits an oscillatory instability, which gives rise
to standing waves. Figures \ref{Fig4} d)-f) show the corresponding
temporarily quasi-periodic states at $\gamma =0.85$ and $\nu =-0.45$, with $%
\lambda _{\max +}=4.92\times 10^{-4}$ and $\lambda _{\max -}=5.13\times
10^{-4}$ (actually, these values may be considered as a numerical zero).
Changing the parameters further in the same direction, the standing wave
exhibits an oscillatory instability, producing complex spatiotemporal
patterns. Figures \ref{Fig4} g)-i) display typical examples of the chaotic
spatiotemporal evolution of the complex patterns, accounted for by $\lambda
_{\max }>0$.

Figure~\ref{Fig6-ArnoldTongue} shows phase diagrams inside the Arnold tongue
for two different initial conditions (see insets): a) a flat profile with a
bump, and b) a small-amplitude periodic pattern. The two respective diagrams
are similar with a few differences, see the comparison between them in panel
c) which displays regions in which both initial conditions produce the same
final state. In the case of the small-amplitude spatially periodic initial
condition, the region of the uniform state (green circles) is somewhat
smaller, and the distribution of regions of temporarily stationary (red
triangles), quasi-periodic (black diamonds), and chaotic (blue squares)
patterns is shifted to smaller values of detuning $\nu $, while
quasi-periodic states appear at larger values of $\gamma $. In both cases,
the dynamical chaos occupies the upper left corner, while near to the
zero-detuning line, $\nu =0$, the uniform state is the dominant one. Thus,
the small-amplitude chaos is equally likely to be generated by both initial
conditions.

%%%%%%%%%%%%%%%%%%%%%%%%%%%%%

\subsection{Dynamics inside the Arnold tongue}

Panel a) of Figure~\ref{Fig6} presents the dependence of $\lambda _{\max }$
on the forcing strength in the case of a spatially periodic pattern chosen as initial condition. In this plot, we observe the transition from a
stationary solution (uniform or patterned one) to an oscillatory one (the
standing wave), which then turns into a spatiotemporal chaotic pattern,
following the increase of the forcing strength, $\gamma $. The growth of $%
\lambda _{\max }$ with the subsequent increase of $\gamma $ indicates that
the chaotic spatiotemporal pattern becomes more complex. Furthermore, around
$\gamma \approx 2.2$, we observe an abrupt transition from $\lambda _{\max
}>0$ to small islands with $\lambda _{\max }=0$, which corresponds to a
phenomenon known as the\textit{\ crisis of a strange attractor} \cite{Ott}.

Further, the patterns are characterized by the plot for $Q_{\max }(\gamma )$
as a function of the forcing strength $\gamma $. Figure~\ref{Fig6} b)
displays it at $\nu =0.1$. It is produced by repeatedly picking the maximum
value of the energy function $Q(\tau )$, see Eq. (\ref{eq7}), from temporal
intervals after the disappearance of transient features in the simulated
evolution. The system is stationary or periodic if there is a unique value
of $Q_{\max }$, while a continuous distribution of $Q_{\max }$ in a finite
interval implies temporal quasi-periodicity or chaos. This figure shows that
energy $Q$ strongly depends on forcing $\gamma $, showing several
transitions between regular and chaotic states, denoted by the same symbols
as in Fig.~\ref{Fig6-ArnoldTongue}. The transitions are shown in greater
detail in the inset, where the plot is blown up by a factor of $5.4$. The
diagrams for $\lambda _{\max }\ $and $Q_{\max }$ provide mutually
complimentary descriptions of the oscillatory, quasi-periodic, and chaotic
behavior. Note that the plot of $Q_{\max }$ is represented by a thin curve
at $\gamma <1.8$, becoming fuzzy at larger values of $\gamma $. In
particular, alternation of islands, from chaos to regularity, is observed
in the range of $1.92<\gamma <2.0$. The system becomes chaotic again at $%
\gamma >2.0$ through an abrupt transition, and remains chaotic at larger
values of $\gamma $ considered here.

To grasp the route to chaos in a clearer form, we have computed the spatial
Fourier spectrum $S_{L}(k)$, as per Eq. (\ref{S}). Figure~\ref{Fig6} shows $%
S_{L}(k)$ for the spatially uniform state at $\gamma =1.50$ in panel c), for
a temporally quasi-periodic one at $\gamma =1.95$ in d), and for a chaotic
state at $\gamma =2.1$ in e). Increase of the number of frequency peaks in
the spectrum is observed as $\gamma $ increases at a fixed value of the
detuning, $\nu =-0.1$. These plots corroborate that Eq. (\ref{eq4}) produces
a standing wave with a well-defined wavenumber and its harmonics in Fig.~\ref%
{Fig6} c). As the forcing strength increases, the system generates
incommensurate wavenumbers, which causes the spectral peaks to spread, see
Fig.~\ref{Fig6} d). Further increase of the strength makes the former sharp
peaks broad in Fig.~\ref{Fig6} e), which is a hallmark of spatial chaos.

\begin{figure}[tbp]
\begin{center}
\includegraphics[width=0.9\textwidth]{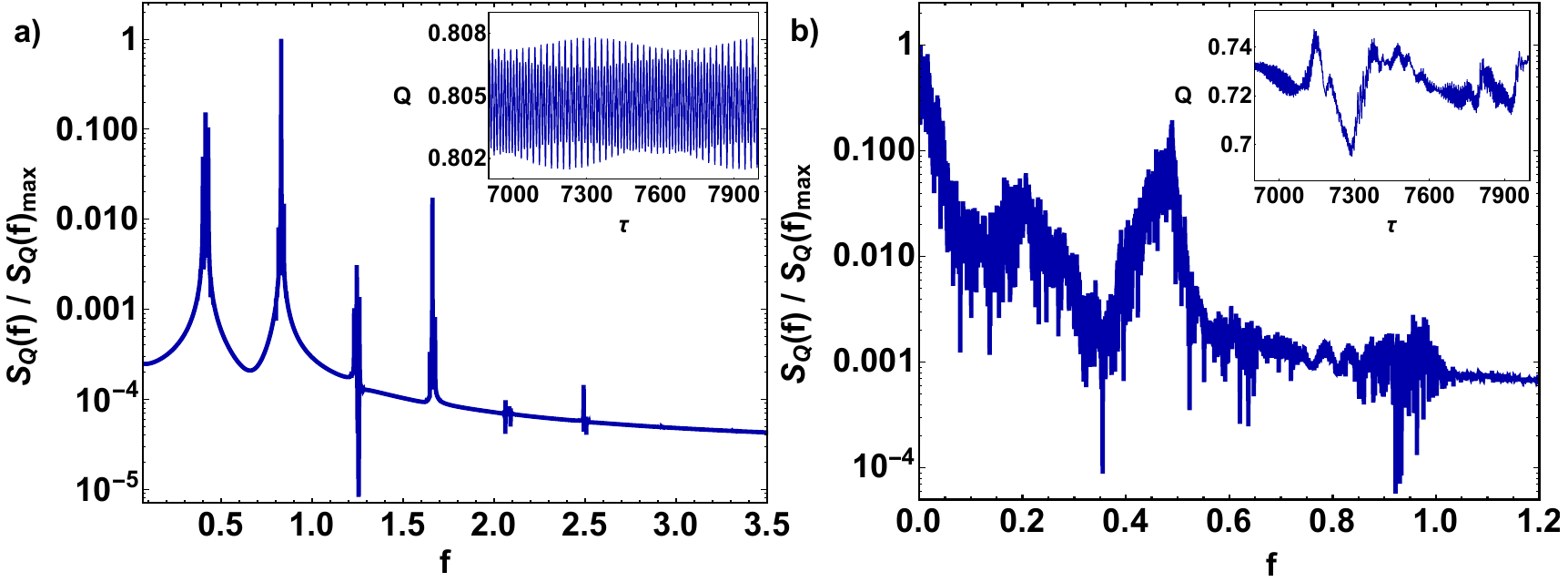}
\end{center}
\caption{The power spectrum $S_{Q}(f)$ of the norm (energy) $Q(t)$, defined
as per Eq. (\protect\ref{eq7}). a) A quasi-periodic state at $\protect\gamma %
=1.89$ and $\protect\nu =-0.1$. b) A chaotic state at $\protect\gamma =0.85$
and $\protect\nu =-0.55$. The insets show the respective plots of $Q(t)$.}
\label{Fig7}
\end{figure}

Further understanding of these patterns is obtained by considering the
temporal evolution of the energy function $Q(t)$, defined by Eq.~(\ref{eq7}%
), and the corresponding Fourier power spectrum, $S_{Q}(f)$. Panel a) of
Fig.~\ref{Fig7} shows $S_{Q}(f)$ for a quasi-periodic pattern at $\protect\gamma =1.89$ and $\protect\nu =-0.1$,
with $\protect\lambda _{\max }=-3.234\times 10^{-4}$. The spectrum features a set of discrete peaks at particular
frequencies. Accordingly, the inset shows regular evolution of $Q(t)$. On
the other hand, panel b) of Fig. \ref{Fig7}  exhibits a chaotic state at $
\protect\gamma =0.85$ and $\protect\nu =-0.55$, with $\protect\lambda _{\max
}=1.288\times 10^{-2}$ showing a continuous spectrum $S_{Q}(f)$,
typical for chaotic states, with irregular evolution of $Q(t)$ in the
respective inset. Thus, the transition to the spatiotemporal chaos proceeds
through an intermediate stage which features the temporal quasi-periodicity.

%%F

%%%%%%%%%%%%%%%%%%%%%%%%%%%
%%%%%%%%%%%%%%%%%%%%%%%%%%%%

\section{Conclusions}

\label{sec:4}

In this work, we have investigated the transition to spatiotemporal chaos
from small-amplitude patterns in damped driven systems close to the
parametric resonance. Starting from the continuum limit of a chain of driven
pendula, which is based on the equation of the sine-Gordon type, the
analysis is performed in the framework of the generalized PDDNLS
(parametrically driven damped nonlinear Schr\"{o}dinger) equation, which is
produced by the small-amplitude expansion of the underlying sine-Gordon
equation. The systematic numerical analysis demonstrates that the onset of
the spatiotemporal chaos proceeds through an intermediate
temporally-quasi-periodic route. We use the generalized PDDNLS equation~(\ref%
{eq4}), which contains higher-order nonlinear terms, because the standard
PDDNLS model, Eq.~(\ref{eq3}), fails to produce stable spatially uniform
solutions, which play a crucial role in obtaining localized states. In
particular, we have found the small-amplitude patterns that are stationary,
quasi-periodic, or chaotic in time depending on the values of the strength
of the parametric forcing, $\gamma $, and resonance detuning, $\nu $. We
have concentrated on chaotic solutions, using different tools to
characterize their dynamical behavior, such as the time dependence of the
norm (energy) $Q$, the corresponding power spectrum, and the maximum
Lyapunov exponent. Small-amplitude chaotic patterns exist inside of the
Arnold tongue, competing with regular ones. Varying $\gamma $ at constant $%
\nu $ or  vice versa leads in various transitions between these states. There
are well confined chaotic patterns and others, in which the chaotic area
penetrates the homogeneous one, following the increase of $\gamma $.

The universal nature of the present model suggests that similar scenarios of
the transition to the small-amplitude spatiotemporal chaos may be expected
in physical settings such as vertically oscillating fluid layers, magnetic
systems, forced nonlinear lattices, and optical waveguides. A promising
direction for the extension of this work is to extend it to the two-dimensional
case.

%%%%%%%%%%%%%%%%%%%%%%%%%%%%%
%%%%%%%%%%%%%%%%%%%%%%%%%%%%%

\section*{Acknowledgements}

LMP and DL appreciate the hospitality of the MPI-P (Mainz) during their stay
in Germany. LMP, PD and DL acknowledge partial financial support from
FONDECYT 1180905. MGC acknowledges partial financial
support from Millennium Institute for Research in Optics, ANID--Millennium
Science Initiative Program--ICN17\_012 and FONDECYT 1180903. JAV and DL acknowledge partial financial
support from Centers of excellence with BASAL/CONICYT financing, Grant
AFB180001, CEDENNA. The work of BAM is supported, in part, by the Israel
Science Foundation through grant No. 1286/17. This author also acknowledges
support from Instituto de Alta Investigaci\'{o}n, Universidad de Tarapac\'{a} (Arica, Chile).

%%%%%%%%%%%%%%%%%%%%%%%%%%%%%
%%%%%%%%%%%%%%%%%%%%%%%%%%%%%

\end{document}